\documentclass[final,1p,times]{elsarticle} % do not change
\usepackage{graphicx} % do not change
\usepackage{lineno} % do not change

\journal{Nuclear Physics A} % do not change
\begin{document} % do not change

\begin{frontmatter} % do not change

%% QM09Author: Boris Tomasik  
%% Title, author and address info here; please do not use footnotes

% Your Title - please insert
\title{The use of Kolmogorov-Smirnov test in event-by-event analysis}

% Principle author, and co-authors - please insert
\author{Boris Tom\'a\v{s}ik$^{a,b}$, Ivan Melo$^c$, Giorgio Torrieri$^d$, Sascha Vogel$^e$, 
Marcus Bleicher$^e$}

% Address - please insert
\address%
{
$^a$ Univerzita Mateja Bela, Tajovsk\'eho 40, 97401 Bansk\'a Bystrica, Slovakia\\
$^b$ Czech Technical University in Prague, FNSPE, B\v{r}ehov\'a 11, 11519 Prague, Czech Republic\\
$^c$ \v{Z}ilinsk\'a Univerzita, Univerzitn\'a 1, 01026 \v{Z}ilina, Slovakia\\
$^d$ FIAS, Goethe-Universit\"at, 
Ruth-Moufang-Str.\ 1, 60438 Frankfurt, Germany\\
$^e$ Institut f\"ur Theoretische Physik, Goethe-Universit\"at,
Max-von-Laue-Str. 1,
60438 Frankfurt, Germany
}

\begin{abstract} % do not change
%% Text of abstract goes here - please insert
We propose to use the Kolmogorov-Smirnov test to uncover non-statistical 
differences between events created in heavy ion collisions within the same 
centrality class. The advantage of the method over other approaches
which are currently in use, is that it is sensitive to any difference between the 
events and is not restricted to simple moments of the distribution of hadrons. 
The particular application examined here is the identification of the fireball 
decay due to spinodal fragmentation and/or sudden rise of the bulk viscosity. 
\end{abstract} % do not change

\end{frontmatter} % do not change

%% QM09: we keep linenumbers at least for initial version
%\linenumbers % do not change

%% start of main text - please insert. 

The hot matter created early in ultrarelativistic heavy ion collisions expands 
very quickly and cools down. At RHIC, data from jet quenching suggest that the system spends a substantial amount 
of time in the deconfined phase \cite{Adams:2003im}. 
The onset of deconfinement is suspected at 
collision energies around $\sqrt{s_{NN}} = 7-8~A$GeV \cite{Seyboth:2005rn}. 
Qualitatively the 
phase diagram of QCD matter shows a smooth though rapid crossover at small baryon 
densities, which are created at RHIC, while a first order phase transition line appears 
at some non-vanishing baryonic chemical potential. 

Generally, the initial  conditions for the fireball and the inner pressure lead to a very 
fast expansion of the fireball. Thus, the matter may pass the phase transition/crossover 
not as slowly, as required for a description in terms of equilibrium
thermodynamics. Such an explosive expansion can lead to non-equilibrium phenomena, like 
supercooling, or even spinodal decomposition. The latter appears in case of a very fast 
expansion through a first-order phase transition when the matter reaches an inflection 
point of the dependence of entropy on some extensive variable. Spinodal decomposition is 
known to happen in nuclear collisions at lower energies where the liquid-gas phase transition is probed
\cite{Chomaz:2003dz}. 
Hence, the fireball decays into smaller fragments which recede from each other and subsequently decay into final state hadrons. 

Such a scenario might seem irrelevant at RHIC as it requires a first order phase 
transition. However, it has been suggested recently that also in the low baryon density region of the 
phase diagramme, fragmentation of the fireball might appear
\cite{Torrieri:2007fb}. 
Here it is due to the sudden increase of the bulk viscosity, which has a sharp peak at the critical temperature $T_c$.
I.e., the expansion starts early in the partonic phase 
and is already very strong when the critical temperature is reached. In this moment
the peak in the bulk viscosity suddenly tries to stop the expansion. As a result of the competition
between inertia and the bulk viscosity the fireball can fragment. 

Thus, the fragmentation phenomenon might be present in nuclear collisions studied 
currently at RHIC.
It is therefore relevant to explore methods which can identify the source break-up mechanism. We realise that fragmentation 
leads to hadronic distributions which will be different in each event. In fact, hadrons are 
produced with velocities close to those of the emitting fragments. Hence, in the distribution 
of hadron momenta from a single event one expects clusters centered around values given 
by the fragment velocities. In each event, these clusters will be at different positions, so there 
will be non-statistical differences between the events even if a sample from a very narrow 
centrality interval is selected. 

Up to now, many techniques have been proposed to investigate the presence of clusters in momentum 
distributions. Among them are rapidity correlations \cite{Pratt:1994ye,rand}, 
correlations in azimuthal angle and pseudorapidity \cite{Alver:2009id}, 
multiplicity fluctuations \cite{Baym:1999up} and mean $p_t$ fluctuations \cite{Baym:1999up,Broniowski:2005ae}.
However, these methods always focus on certain \emph{moments} of the momentum distribution. 
In contrast to this, here we propose a method which compares the complete event \emph{shapes}. 

We use the Kolmogorov-Smirnov (two-sample two-sided) test (KS test), 
which can be used to measure the 
similarity between two empirical sets of data \cite{kolm,smir}. 
It answers the question, to what extent two 
sets of data are generated by the same mechanism with the same underlying probability density.
To compare two events, one first constructs the empirical cumulative distribution function 
for each event. We use here the measured rapidities of hadrons. On the abscissa we put
all the measured rapidities in one event. Then we draw a ``staircase'' by putting at each of the 
positions of rapidities a step of the height $1/n$, where $n$ is the multiplicity of 
the event. This is done 
for two events and the maximum vertical
distance between the staircases (denoted $D$) is taken as the 
measure of the difference between the events. For large multiplicities, the cumulative 
distribution function of the quantity $\sqrt{n}D$ is known, provided that the events are 
generated from the same distribution 
\begin{equation}
\label{cdf}
P(\sqrt{n} D) := 1 - Q(\sqrt{n}D) 
= \sum_{k=-\infty}^{\infty} (-1)^k \exp (-2 k^2 n D^2) + O(n^{-1/2})\, ,
\qquad
n = \frac{n_in_j}{n_i+n_j} 
\, .
\end{equation}
The variable $Q$ acquires values between 0 and 1.

Let us now explain the method in more detail. 
We start from a large sample of events. For each pair of events we measure $D$ and determine 
$Q$ from equation (\ref{cdf}). (Details of the precise calculation are shown in 
\cite{Melo:2009xh}.)
If all events correspond to the same underlying probability distribution, then the differences 
between them are only statistical and we obtain $Q$'s which are distributed \emph{uniformly}. 
Any departure from the uniform distribution indicates that some of the following assumptions are not fulfilled: 
i) within one event all particles are produced \emph{independently} from each other;  
ii) all events are generated from the \emph{same} probability distribution. 

%%%%%%%%%%%%%%%%%%%%%%%%%%%%%%%%%%%%%%%%%%%%%%%%%%

The advantage of the method is that it is sensitive to any effects breaking the two 
assumptions. To demonstrate the power of the method, we provide an example of data where two-particle 
correlations show nothing but the KS test gives a non-trivial result. 
We generate \emph{toy events} with a mean multiplicity of 1000. Particles have the same 
mass and are distributed into 100 groups. Hence, on average there are ten particles
per group. Their momenta are generated from a uniform and spherical 
distribution. The momentum of the last particle in each group is calculated 
so  that the total momentum within the group vanishes.  
%%%%%%%%%%%%%%%%%%%%%%%%%%%%%%%%%%%%%%%%%%%%%%%%%%%%%%%%%%%%%%%%%%%
\begin{figure}[ht]
\centering
\includegraphics[width=0.49\textwidth]{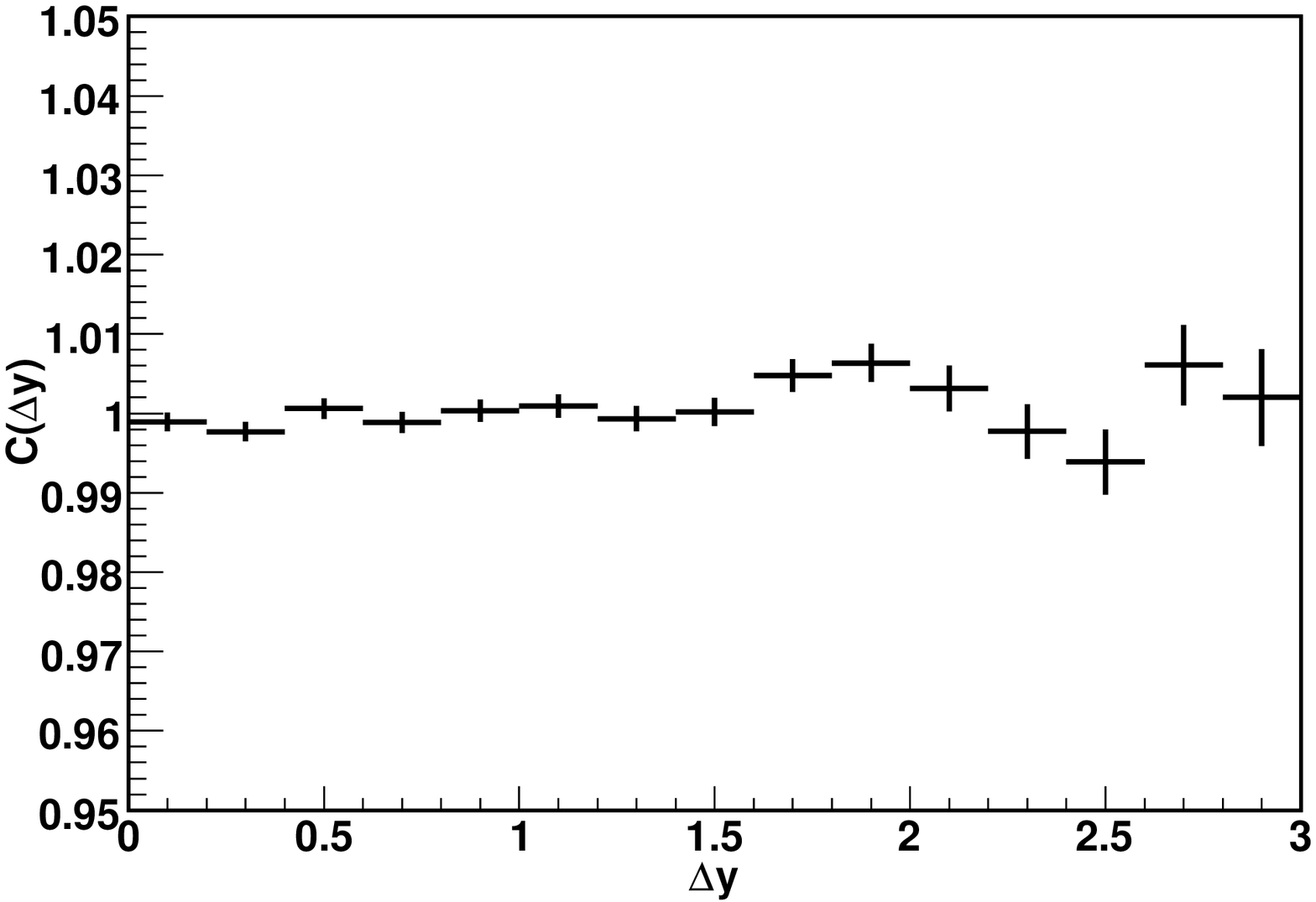}
\includegraphics[width=0.49\textwidth]{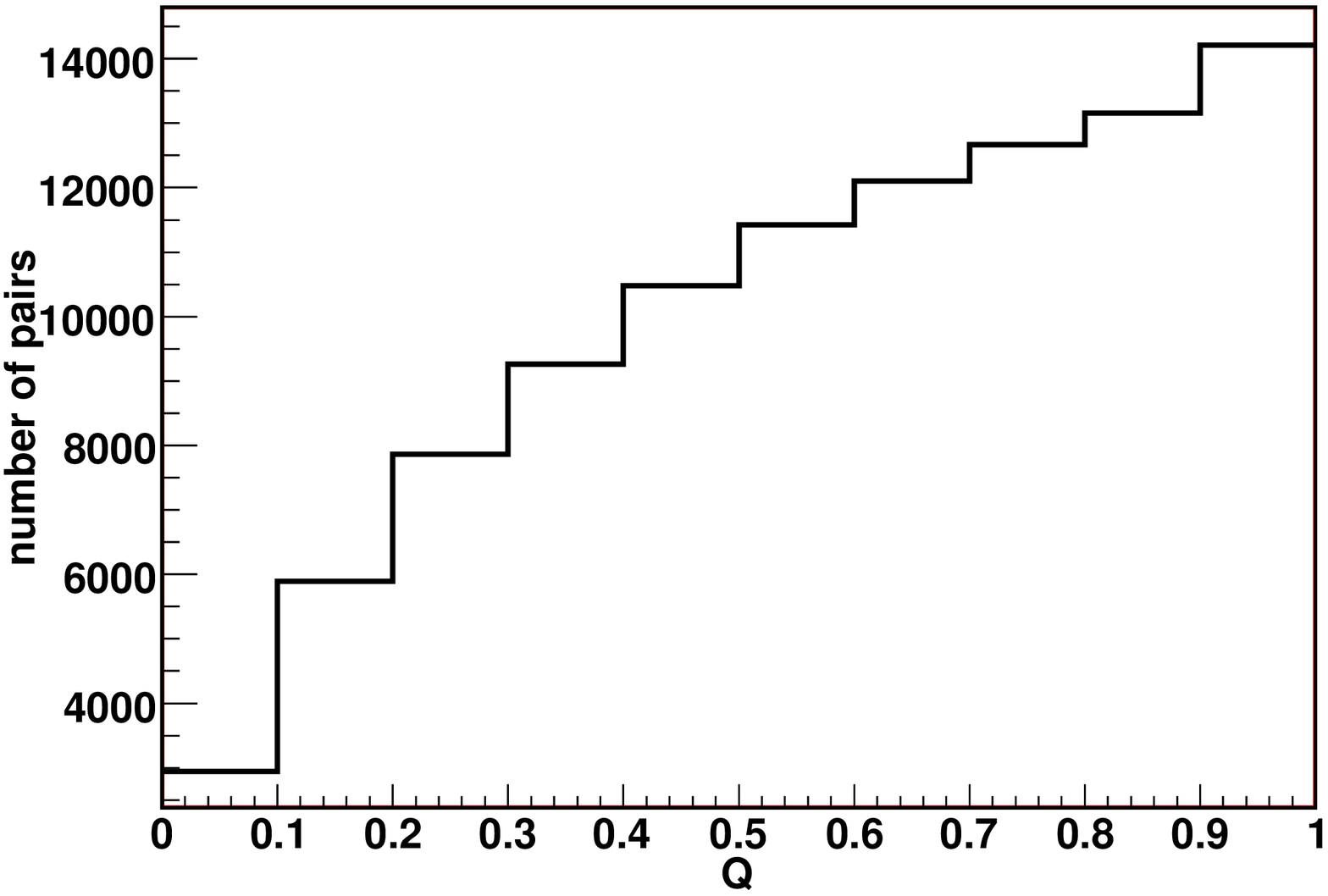}
%%
%% here is an example syntax if you want to have a single eps file in the figure
%\includegraphics[scale=0.19]{integrated_charm_with_FONLL.eps}	       
%% - then you need to use commands a la those shown in the included 'mkpaper'
%% script: e.g. by doing 'source mkpaper'
%%
%% next is an example for including 2 non-eps files in the same figure
%\includegraphics[width=0.4\textwidth]{integrated_charm_with_FONLL.jpg}
%\includegraphics[width=0.45\textwidth]{all_spectras_together.jpg}
%% in the example here we give slightly more horiz. space for the 2nd figure 
%% - then you can just use 'pdflatex example.tex', and get a pdf directly
%%
\caption[]{
(Left) Two-particle correlation function in rapidity difference between particles 
generated by the toy generator with simple momentum conservation. (Right) The 
$Q$-histogram of 10$^5$ pairs of events from the same data.
}
\label{f:toy}
\end{figure}
%%%%%%%%%%%%%%%%%%%%%%%%%%%%%%%%%%%%%%%%%%%%%%%%%%%%%%%%%%%%%%%%%%%%%%
In this construction, there is  no two-particle 
correlation between the particles, as presented in Figure~\ref{f:toy}. (Note that 
a ten-particle correlation would show a signal.) The 
KS test, however, shows very clear deviations from a uniform distribution 
(Fig.~\ref{f:toy}). 
To quantify the deviation, we introduce the parameter 
\begin{equation}
\label{Rdef}
R
= \frac{N_0 - \frac{N_{\rm tot}}{B}}{\sqrt{\frac{N_{\rm tot}}{B}}}\, ,
\end{equation}
where $N_0$ is the number of pairs in the first bin, 
$N_{\rm tot}$ is the total number of pairs, and
$B$ is the number of bins of the $Q$-histogram. Momentum conservation 
leads to negative $R$. Later we show that fireball fragmentation 
causes large positive $R$.

%%%%%%%%%%%%%%%%%%%%%%%%%%%%%%%%%%%%%%%%%%%%%%%%%%%%%%%

To test the method on realistic data, we generate event samples with the Monte Carlo 
event generator DRAGON \cite{Tomasik:2008fq}. DRAGON generates the momenta of 
hadrons as if they were produced from a fragmented fireball. The pattern 
of expansion is that of the blast-wave model. Here, the blast wave model is used 
to generate the positions and velocities of the fragments which subsequently 
radiate hadrons. Some hadrons can be produced also from the space between 
the fragments. The chemical composition is determined according to the 
grand-canonical ensemble and resonance decays are taken into account. Choosing 
different values for the parameters of the model (temperature, fireball sizes,
droplet sizes, etc.) allows then to simulate different physical situations. 

With DRAGON, sets of $10^4$ events are generated out of which we randomly choose 
$10^5$ pairs. On these pairs we evaluate the variable $Q$ and fill histograms. 
As mentioned above, a departure from a flat distribution indicates 
non-statistical differences between the events or that the particles within 
one event are not emitted independently from each other. 

%%%%%%%%%%%%%%%%%%%%%%%%%%%%%%%%%%%%%%%%%%%%%%%%%%%%%%%%%%%%%%%%%%%
\begin{figure}[th]
\centering
\includegraphics[width=0.49\textwidth]{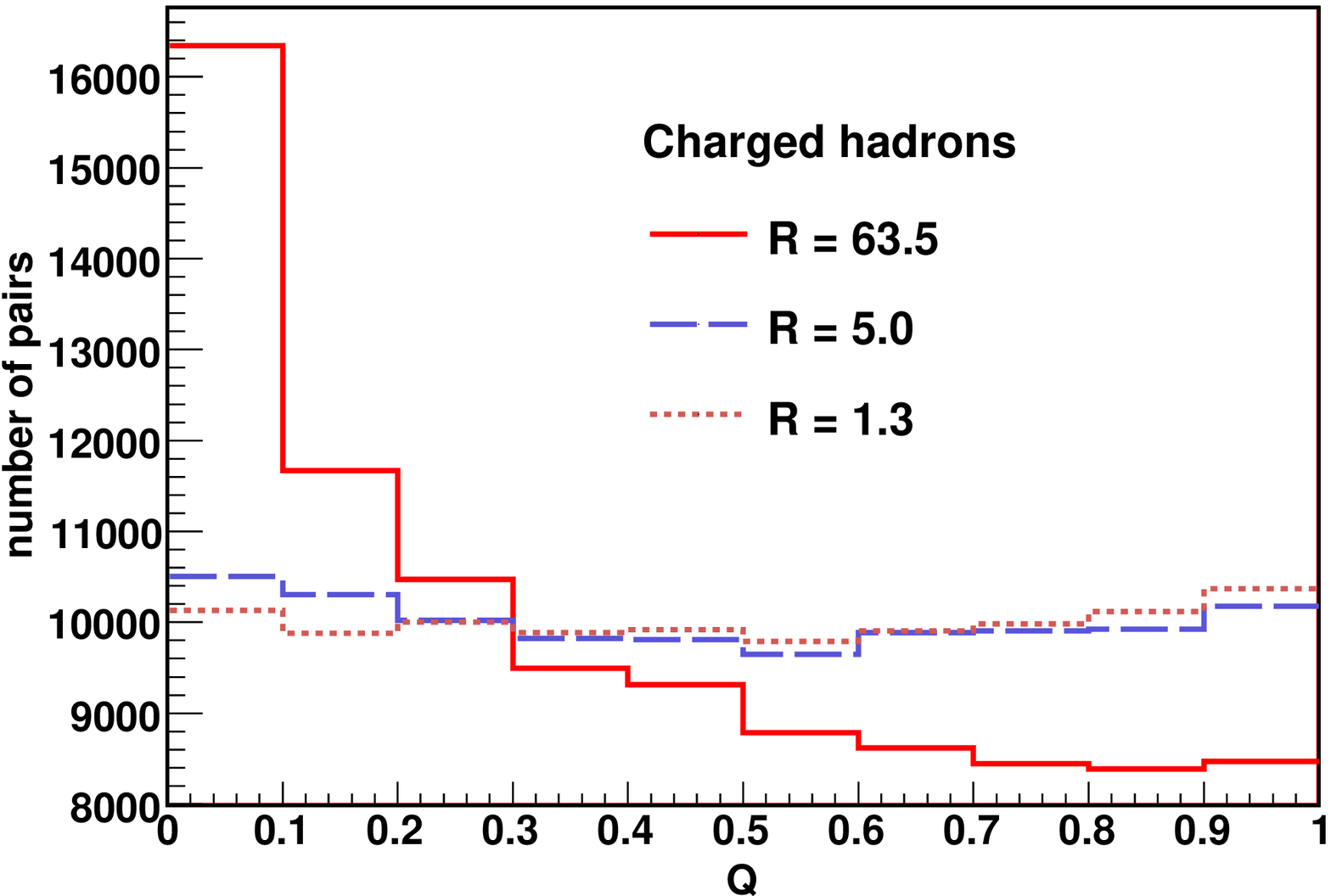}
\includegraphics[width=0.49\textwidth]{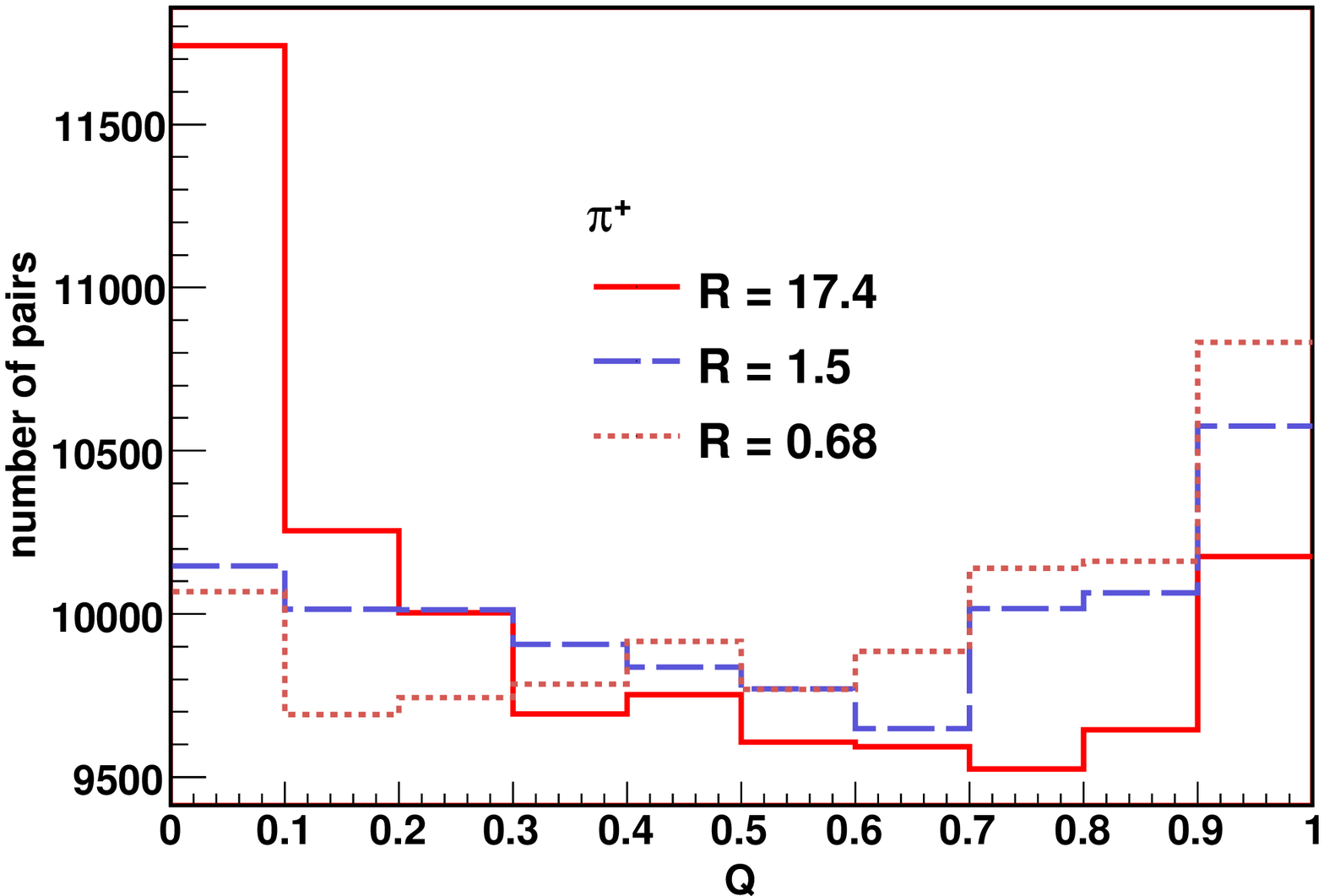}
%%
%% here is an example syntax if you want to have a single eps file in the figure
%\includegraphics[scale=0.19]{integrated_charm_with_FONLL.eps}	       
%% - then you need to use commands a la those shown in the included 'mkpaper'
%% script: e.g. by doing 'source mkpaper'
%%
%% next is an example for including 2 non-eps files in the same figure
%\includegraphics[width=0.4\textwidth]{integrated_charm_with_FONLL.jpg}
%\includegraphics[width=0.45\textwidth]{all_spectras_together.jpg}
%% in the example here we give slightly more horiz. space for the 2nd figure 
%% - then you can just use 'pdflatex example.tex', and get a pdf directly
%%
\caption[]{$Q$-histograms from a simulation for central Au+Au reactions at $\sqrt{s_{NN}}=200$~GeV (RHIC) 
with assumed fragmentation 
into fragments with mean volume of 5~fm$^3$ (solid red lines). For comparison, 
results from a non-fragmented fireball at RHIC (dashed blue lines) and FAIR ($E_{beam}=30A$GeV)
(dotted brown lines) are shown. Left: rapidities of all charged hadrons are used for 
the KS test. Right: only rapidities of charged pions are used.}
\label{fig1}
\end{figure}
%%%%%%%%%%%%%%%%%%%%%%%%%%%%%%%%%%%%%%%%%%%%%%%%%%%%%%%%%%%%%%%%%%%%%%
In Figure~\ref{fig1} we show how the fragmentation of the fireball is reflected in 
$Q$-histograms. Results from a simulation with fragments with an average volume of
5~fm$^3$ are compared to two cases with direct emission of hadrons. Parameters in the simulation
with fragments are chosen to correspond to Au+Au collisions at RHIC. The KS test 
is performed with the rapidities of the hadrons. In the histogram obtained 
for all charged hadrons one clearly observes that fragmentation leads to a pronounced peak 
at small $Q$. 
For comparison, simulations without fragments (i.e. the hadrons are emitted directly from the source) are made and studied. Naively, one would expect flat $Q$-histograms in these cases. However, one observes that 
it is not the case (Fig.~\ref{fig1}, left). For charged hadrons, there are correlations 
between the produced hadrons due to resonance decays, which actually act
as small clusters. This (unwanted) correlation can be minimized by using only pions of one charge, 
as shown in the right panel. Indeed, the peak at small $Q$ disappears (note the different scales). 
However, we have decreased the multiplicity of hadrons entering the 
procedure and this leads to a peak close to $Q=0$ due to 
applicability limits of the approximative
formula (\ref{cdf}) (even though we actually use an improved formula). 

We have also checked that bigger droplets and a larger abundance of hadrons 
emitted from the droplets leads to a more pronounced low-$Q$ peak. 

%%%%%%%%%%%%%%%%%%%%%%%%%%%%%%%%%%%%%%%%%%%

In summary, analyzing data with the Kolmogorov-Smirnov test can 
provide novel insights and identify interesting effects not seen
before with other methods. Here, the proposed particular use 
was for the identification of a possible fragmentation of the 
fireball via spinodal decomposition or due to 
sudden appearance of the bulk viscosity. 
We showed that such a fragmentation can be clearly identified  
with the help of the KS technique. While momentum and charge conservation 
yield only minor modifications, the influence of other processes 
needs yet to be studied. It is often assumed that in narrow centrality 
interval all events develop according to the same physics scenario. We 
propose to test this assumption with the method described here.

%% end of main text
%%%%%%%%%%%%%%%%%%%%%%%%%%%%%%%%%%%%%%%%%%%%%%%%%%%%%%%%%%%%%%%%%%%

\section*{Acknowledgments} % please insert, comment out or delete if not needed
BT and IM thank for support by VEGA 1/4012/07 (Slovakia). BT also acknowledges support from 
MSM~6840770039 and  LC~07048 (Czech Republic). The work of GT, SV, and MB 
was supported by the Helmholtz International
Center for FAIR within the framework of the LOEWE program
(Landesoffensive zur Entwicklung Wissenschaftlich-\"okonomischer
Exzellenz) launched by the State of Hesse. The collaboration between 
the Slovak and the German group was facilitated by funding from DAAD.

%This is where one places acknowledgments for funding bodies etc., if needed.
%For the large collaborations, this is listed once and for all, together with 
%the author lists etc. in the proceedings back-material.

%%%%%%%%%%%%%%%%%%%%%%%%%%%%%%%%%%%%%%%%%%%%%%%%%%%%%%%%%%%%%%%%%

 % do not change 
\end{document}